\documentclass[aps,prd,preprint,12pt,superscriptaddress,nofootinbib,a4paper]{revtex4-1}

\usepackage[utf8]{inputenc}
\usepackage{amsmath,amssymb}
\usepackage[separate-uncertainty=true]{siunitx}
\usepackage{graphicx}
\usepackage[usenames,dvipsnames]{xcolor}
\usepackage[caption=false]{subfig}
\usepackage{hyperref}
\usepackage[capitalize]{cleveref}
\usepackage{booktabs}
\usepackage{tabularx}
\usepackage{xspace}
\usepackage{color}
\usepackage[normalem]{ulem}
\usepackage{slashed}
\usepackage{bbold}
\usepackage{wasysym}
\usepackage{graphicx}
\usepackage{mathrsfs} 

\allowdisplaybreaks

\graphicspath{{graphics/}}

\newcommand{\eqn}{equation}
\newcommand{\lb}{\left(}
\newcommand{\rb}{\right)}
\newcommand{\be}{\beta}
\newcommand{\al}{\alpha}

\newcommand{\GeV}{{\ensuremath\rm GeV}}

\newcommand{\pb}{{\ensuremath\rm pb}}
\newcommand{\fb}{{\ensuremath\rm fb}}
\newcommand{\ab}{{\ensuremath\rm ab}}

\DeclareSIUnit{\pb}{pb}
\DeclareSIUnit{\fb}{fb}
\AtBeginDocument{
\heavyrulewidth=.08em
\lightrulewidth=.05em
\cmidrulewidth=.03em
\belowrulesep=.65ex
\belowbottomsep=0pt
\aboverulesep=.4ex
\abovetopsep=0pt
\cmidrulesep=\doublerulesep
\cmidrulekern=.5em
\defaultaddspace=.5em

\newcolumntype{C}{>{\centering\arraybackslash}X}
\newcolumntype{b}{C}
\newcolumntype{s}{>{\hsize=.6\hsize}C}
\newcolumntype{R}{>{\raggedleft\arraybackslash}X}
}

\begin{document}
\pdfoutput=1

\bibliographystyle{hunsrt}
\date{\today}
\rightline{RBI-ThPhys-2022-10, CERN-TH-2022-042}
\title{{\Large A short overview on low mass scalars at future lepton colliders - Snowmass White Paper}}

\author{Tania Robens}
\email{trobens@irb.hr}
\affiliation{Ruder Boskovic Institute, Bijenicka cesta 54, 10000 Zagreb, Croatia}
\affiliation{Theoretical Physics Department, CERN, 1211 Geneva 23, Switzerland}

\renewcommand{\abstractname}{\texorpdfstring{\vspace{0.5cm}}{} Abstract}

\begin{abstract}
    \vspace{0.5cm}
  In this whitepaper, I give a short summary on possible channels of low-mass scenarios and their discovery potential at future $e^+e^-$ colliders. This is a summary of talks I recently gave at the CEPC workshop, FCC week and ECFA future collider workshop.
\end{abstract}

\maketitle

\section{Introduction}
The discovery of a scalar which so far largely agrees with predictions for the Higgs boson of the Standard Model (SM) has by now been established by the LHC experiments, with analyses of Run II LHC data further confirming this. In the European Strategy Report \cite{EuropeanStrategyforParticlePhysicsPreparatoryGroup:2019qin,European:2720129}, a large focus was put on future $e^+e^-$ colliders, especially so-called Higgs factories with center-of-mass (com) energies around $240\,-250\,\GeV$. While these will on the on hand further help to determine properties of the scalar discovered at the LHC, and especially will help to determine in detail the parameters and shape of the scalar potential, it is also interesting to investigate theis potential to search for additional scalar states. Many new physics models still allow for extra scalar states, including those which have masses $\lesssim\,125\,\GeV$.\\

In this whitepaper, I give a short summary of presentations I gave at various recent meetings and workshops. I give an overview on some models that allow for such light states, and point to phenomenological studies investigating such models. This should be viewed as an encouragement for further detailed studies in this direction.
\section{Models}
\subsection{Singlet extensions}
In singlet extensions, the SM scalar potential is enhanced by additional scalar states that are singlets under the SM gauge group. In such scenarios, the coupling of the novel scalar to SM particles is typically inherited via mixing, i.e. mass-eigenstates are related to gauge eigenstates via a unitary mixing matrix. The corresponding couplings and interactions are mediated via a simple mixing angle.

\begin{figure}
\includegraphics[width=0.75\textwidth]{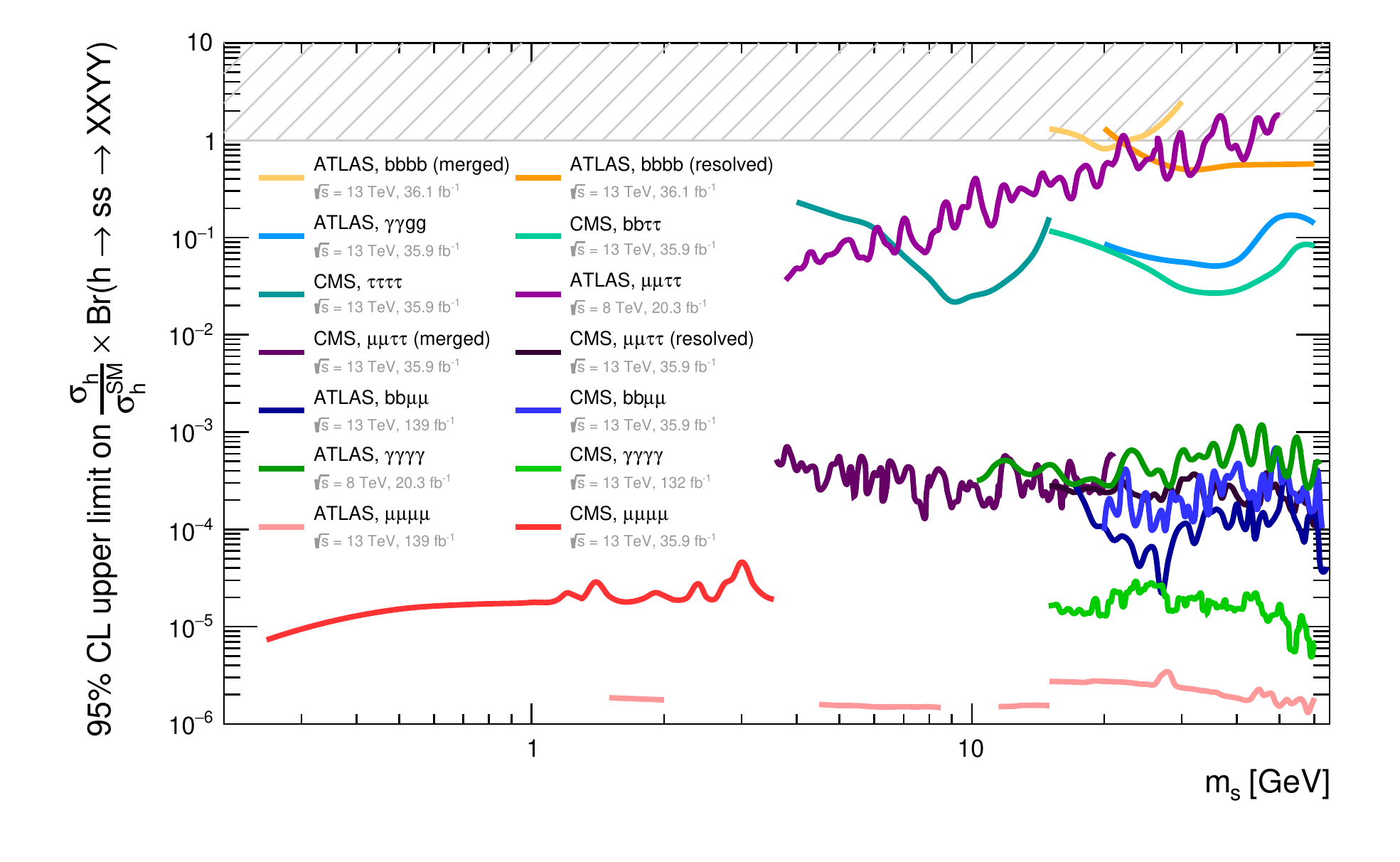}
\caption{\label{fig:current_stat} Limits on the process in eqn (\ref{eq:lims}, taken from \cite{Cepeda:2021rql}. This displays current constraints which can especially be easily reinterpreted in extended scalar sector models, in particular models where couplings are inherited via a simple mixing angle.}
\end{figure}

In \cite{Cepeda:2021rql}, the authors present the status of current searches for the process
\begin{\eqn}\label{eq:lims}
p\,p\,\rightarrow\,h_{125}\,\rightarrow\,s\,s\,\rightarrow\,X\,X\,Y\,Y
\end{\eqn}
which for such models can be read as a bound in 
\begin{\eqn*}
\sin^2\,\theta\,\times\,\text{BR}_{h_{125}\,\rightarrow\,s\,s\,\rightarrow\,X\,X\,Y\,Y}.
\end{\eqn*}
We display these results in figure \ref{fig:current_stat}.

\begin{center}
\begin{figure}
\includegraphics[width=0.45\textwidth]{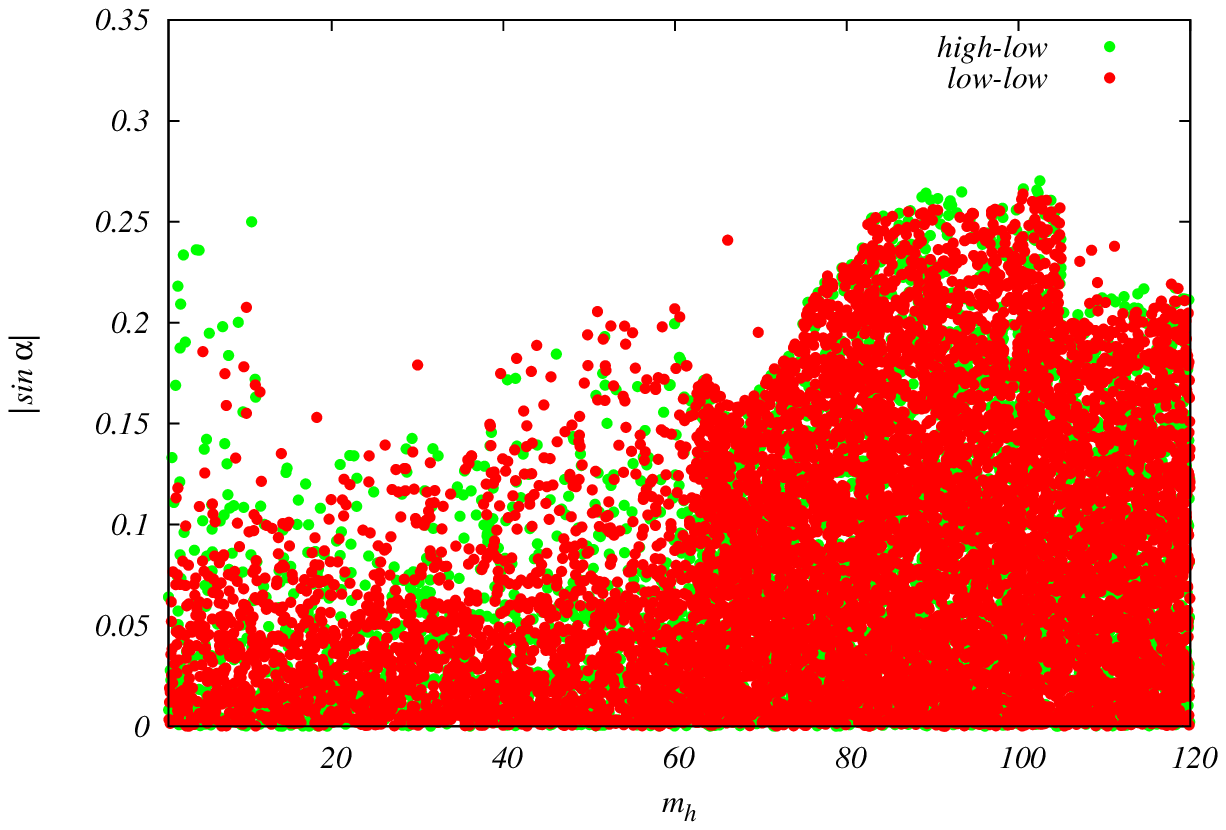}
\includegraphics[width=0.45\textwidth]{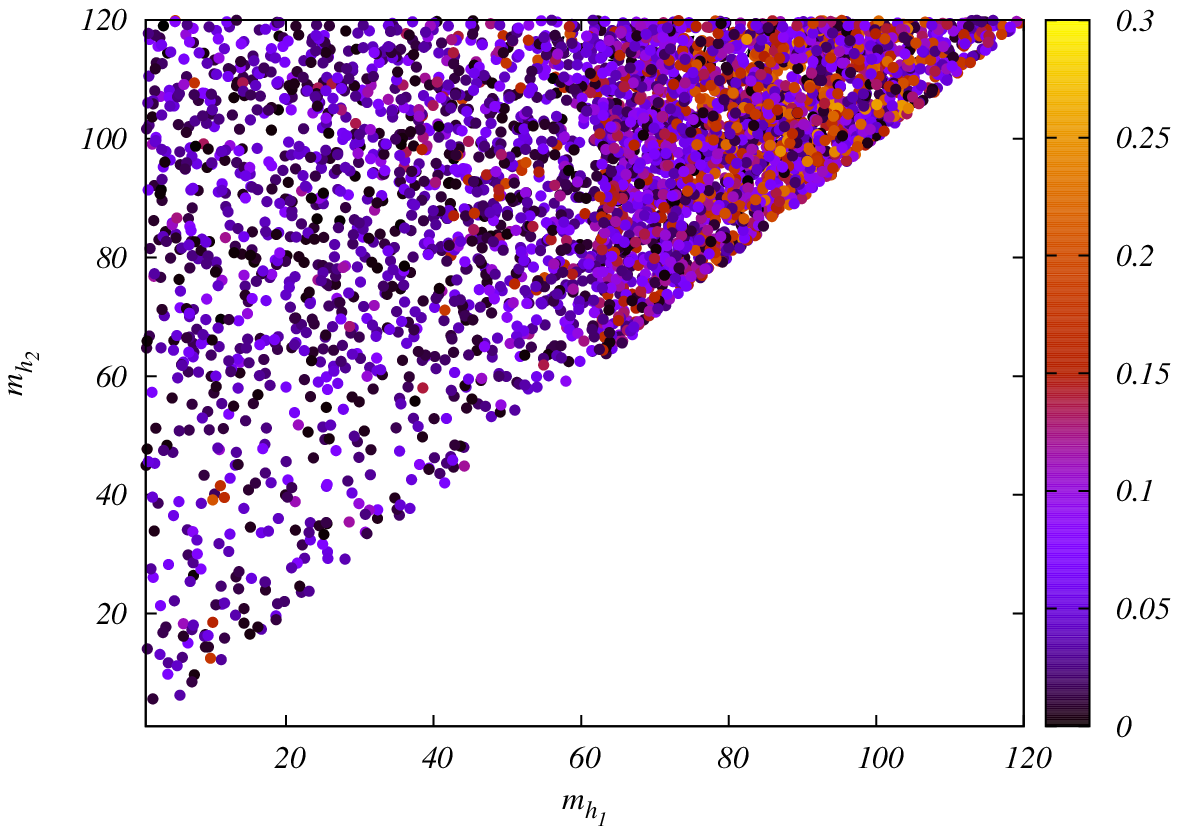}
\caption{\label{fig:trsm} Available parameter space in the TRSM, with one (high-low) or two (low-low) masses lighter than 125 \GeV. {\sl Left}: light scalar mass and mixing angle, with $\sin\al\,=\,0$ corresponding to complete decoupling. {\sl Right:} available parameter space in the $\lb m_{h_1},\,m_{h_2}\rb$ plane, with color coding denoting the rescaling parameter $\sin\al$ for the lighter scalar $h_1$.}
\end{figure}
\end{center}

We show an example of the allowed parameter space in a model with two additional singlets, the two real scalar extension studied in \cite{Robens:2019kga}. In this model, three CP-even neutral scalars exist that relate the gauge and mass eigenstates $h_{1,2,3}$ via mixing. One of these states has to have couplings and mass complying with current measurements of the SM-like scalar, the other two can have higher or lower masses. In figure \ref{fig:trsm}, we show two cases where either one (high-low) or two (low-low) scalar masses are smaller than $125\,\GeV$. On the y-axis, the respective mixing angle is shown. Complete decoupling would be designated by $\sin\al\,=\,0$ in the notation used in this figure.
The points were generated using ScannerS \cite{Coimbra:2013qq,Muhlleitner:2020wwk}, interfaced to HiggsBounds-5.10.2 \cite{Bechtle:2008jh,Bechtle:2011sb,Bechtle:2013wla,Bechtle:2020pkv} and HiggsSignals-2.6.2 \cite{Bechtle:2013xfa,Bechtle:2020uwn}, with constraints as implemented in these versions.
\subsection{Two Higgs Doublet Models}
Two Higgs doublet models (2HDMs) constitute another example of new physics models allowing for low mass scalar states. A general discussion of such models is e.g. given in \cite{Branco:2011iw} and will not be repeated here. In general, such models contain, besides the SM candidate, two additional neutral scalars which differ in CP properties as well as a charged scalar, so the particle content is given by $h,\,H,\,A,\,H^\pm$, where one of the two CP-even neutral scalars $h,\,H$ needs to be identified with the 125 \GeV~ resonance discovered at the LHC. Couplings to the fermions in the Yukawa sector distinguish different types of 2HDMs.

\begin{figure}
\includegraphics[width=0.95\textwidth]{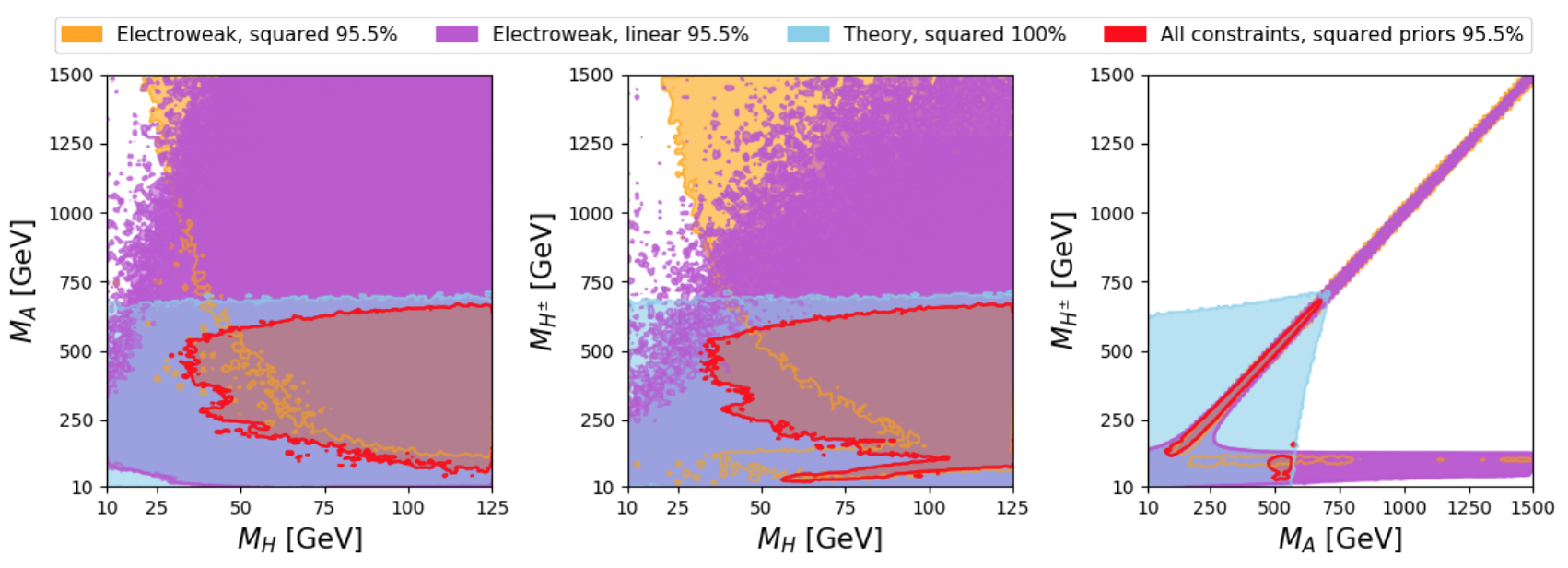}
\caption{\label{fig:2hdm} Allowed regions in the 2HDM, from a scan presented in \cite{Eberhardt:2020dat}.}
\end{figure}

In \cite{Eberhardt:2020dat}, the authors perform a scan including bounds from theory, experimental searches and constraints, as e.g. electroweak observables, as well as B-physics. Examples for these scan results are shown in figure \ref{fig:2hdm}, taken from that reference.
We see that for all regions solutions for either one or several low mass scenarios exist and are viable for the constraints discussed in that reference. Unfortunately, the information about the $\cos\lb \be-\al\rb$ regions in these scenarios is not available in that reference. Depending on the Yukawa couplings considered, the limits on the absolute value of this rescaling angle vary between 0.05 and 0.25 \cite{ATLAS-CONF-2021-053}.
\subsection{Other extensions}
The scalar sector of the SM can be extended by an arbitrary number of additional scalar fields, such as singlets, doublets, etc. One option which is also often consider is the extension of this sector by both singlets and doublets.
\subsubsection{N2HDM} 
\begin{center}
\begin{figure}
\includegraphics[width=0.65\textwidth]{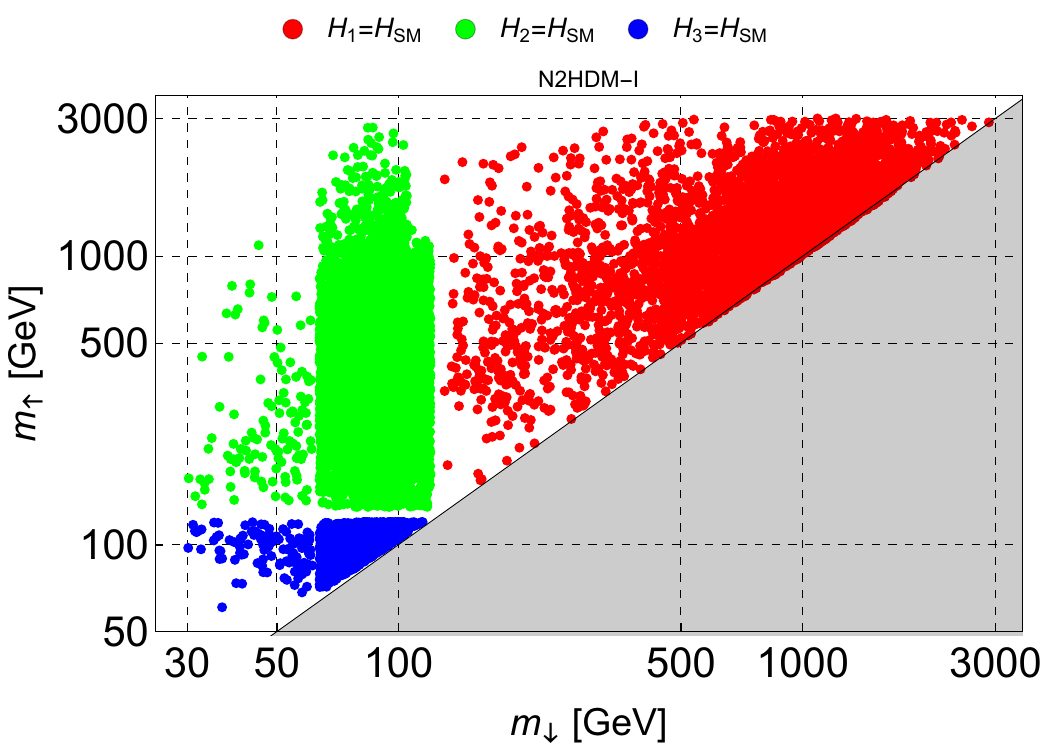}
\caption{\label{fig:n2hdm} Scan results in the N2HDM, taken from \cite{Abouabid:2021yvw}. There are regions in the models parameter space where either one or two of the additional scalars have masses $\lesssim\,125\,\GeV$.}
\end{figure}
\end{center}

In \cite{Abouabid:2021yvw}, the authors consider a model where the SM scalar sector is extended by an additional doublet as well as a real singlet. This model has 3 CP even neutral scalar particles, out of which one needs to have the properties in compliance with LHC measurements of the 125 \GeV~ scalar. The authors perform an extensive scan and find regions in parameter space where either one or both of the additional scalars have masses below 125 \GeV. We show an example of the allowed parameter space in figure \ref{fig:n2hdm}
\subsubsection{Lepton-specific IDM}
\begin{center}
\begin{figure}
\includegraphics[width=0.85\textwidth]{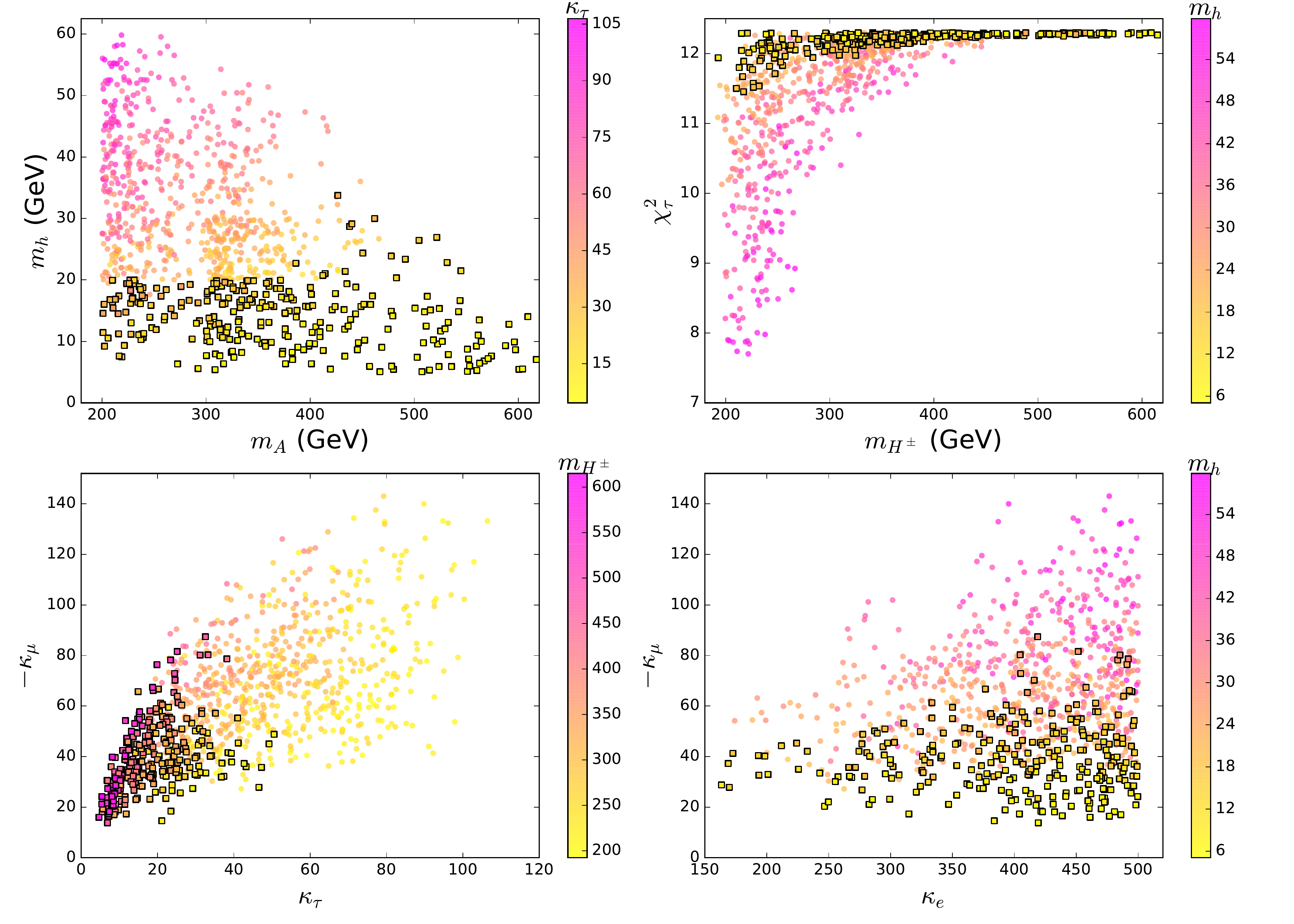}
\caption{\label{fig:idmv} Allowed regions in the parameter space of the model discussed in \cite{Han:2021gfu}, taken from that reference, where squares denote allowed and bullets excluded regions in the models parameter space. CP-even neutral scalars with low masses are viable within this model.}
\end{figure}
\end{center}

In \cite{Han:2021gfu}, the authors consider a model where the SM scalar sector is augmented by an additional doublet, where they impose an exact $\mathbb{Z}_2$ symmetry. This symmetry is then broken by a specific coupling to the fermionic sector. The authors identify regions in the models parameter space that agree with current searches as well as anomalous magnetic momenta of electron and muon. They identify regions in the models parameter space where the second CP-even scalar can have a mass $\lesssim\,30\,\GeV$. We display these regions in figure \ref{fig:idmv}.
\subsubsection{Scalar triplet model}
\begin{center}
\begin{figure}
\includegraphics[width=0.5\textwidth]{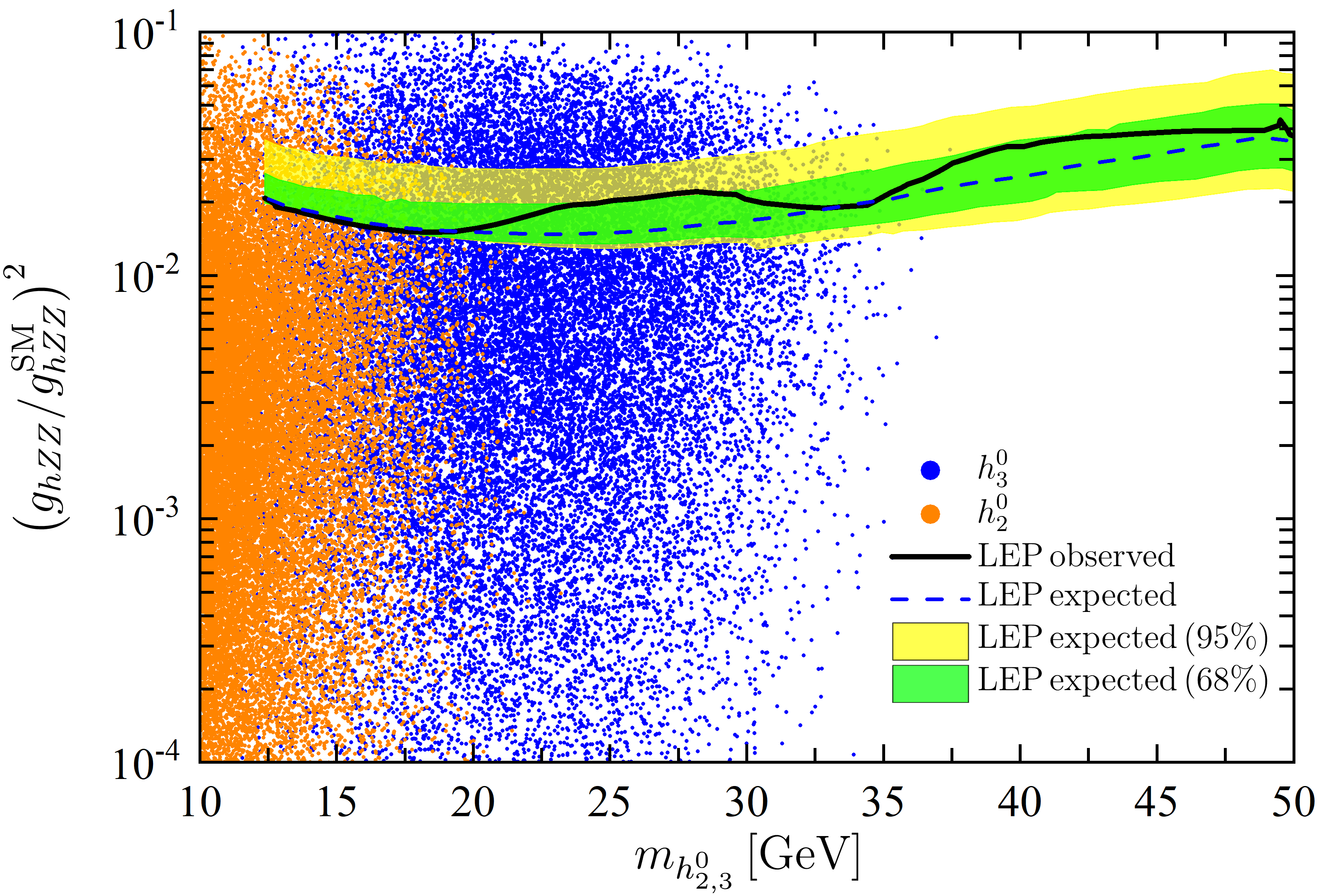}
\includegraphics[width=0.45\textwidth]{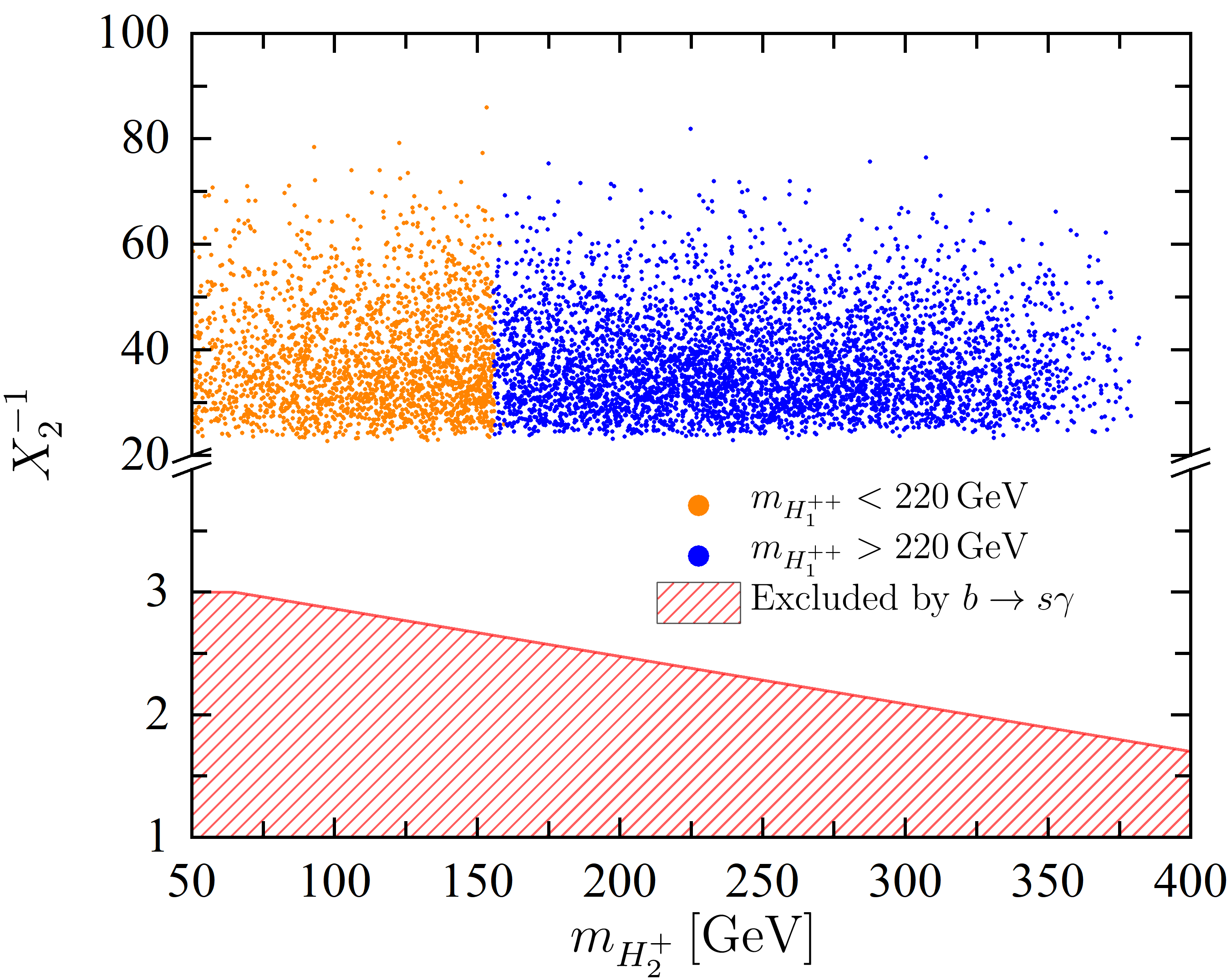}
\caption{\label{fig:tripl} Allowed regions in the parameter space of the model discussed in \cite{Ferreira:2021bdj}, taken from that reference. For neutral and charged new scalars, masses $\lesssim\,125\,\GeV$ are achievable.}
\end{figure}
\end{center}

Finally, we want to discuss a model containing scalar triplets, leading to a rich particle content as well as the possibility of CP violating terms. The model has been presented in \cite{Ferreira:2021bdj}. This model contains 5 neutral, 3 charged, and 2 doubly charged mass eigenstates. The authors present regions in parameter space where masses for some of these can be $\lesssim\,125\,\GeV$. We display these results in figure \ref{fig:tripl}.
\section{Studies at 90 \GeV}

For this center-of-mass energy, several searches exist which have already been performed at LEP and are summarized in \cite{OPAL:2002ifx,ALEPH:2006tnd}, concentrating on $Z\,h$, $h_1\,h_2,$ and $h_1\,h_1\,h_1$ final states, where $h_i$ signifies novel scalars. Possible new studies could build upon these searches. We want to note that the luminosity at FCC-ee and CEPC at this center of mass energy is exceeding LEP luminosity by several orders of magnitude \cite{FCC:2018evy,CEPCStudyGroup:2018ghi}.

\begin{center}
\begin{figure}
\includegraphics[width=0.5\textwidth]{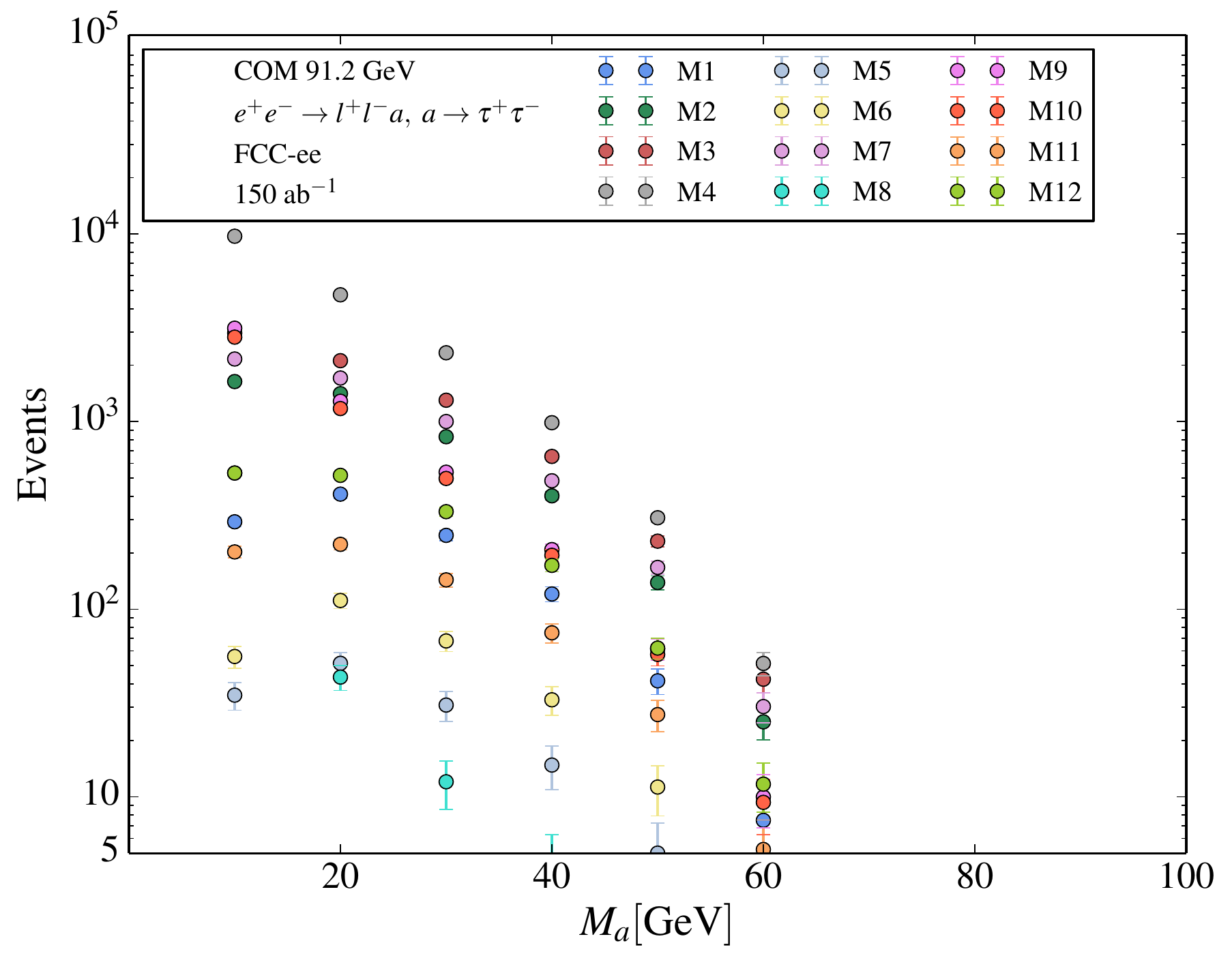}
\caption{\label{fig:benjpred} Rates at a 91 \GeV FCC-ee for various models discussed in \cite{Cornell:2020usb}, for $\ell^+\ell^-\tau^+\tau^-$ final states. M7 and M10 reach a 3 $\sigma$ significance using ML techniques. Figure taken from \cite{Cornell:2020usb}.}
\end{figure}
\end{center}

We also want to present one specific study which investigates several composite models at a com energy of 91 \GeV~ \cite{Cornell:2020usb}. The authors consider the process $e^+e^-\,\rightarrow\,\ell^+\ell^-\tau^+\tau^-$, where the tau-pair stems from an additional pseudoscalar $a$ radiated off one of the fermion lines in the $\ell\ell$ pair-production. They apply a cut-based study as well as an improved analysis using machine learning techniques; for the latter, the authors are able to achieve a 3 $\sigma$ exclusion for benchmarks with masses $M_a\,\sim\,20\,\GeV$. We display event rates for the various benchmark scenarios in figure \ref{fig:benjpred}.

\section{Studies at 240-250 \GeV}
Throughout this work, we show for reference leading order predictions for $Zh$ production at $e^+e^-$ colliders for low mass scalars which are SM-like. These results were obtained using Madgraph5 \cite{Alwall:2011uj} and are given for approximate reference only. We also display the VBF-type production of $e^+e^-\,\rightarrow\,h\,\nu_\ell\,\bar{\nu}_\ell$. Note that the latter signature also contains contributions from $Z\,h$ production, where $Z\,\rightarrow\,\nu_\ell\,\bar{\nu}_\ell$.

\begin{center}
\begin{figure}
\includegraphics[width=0.45\textwidth]{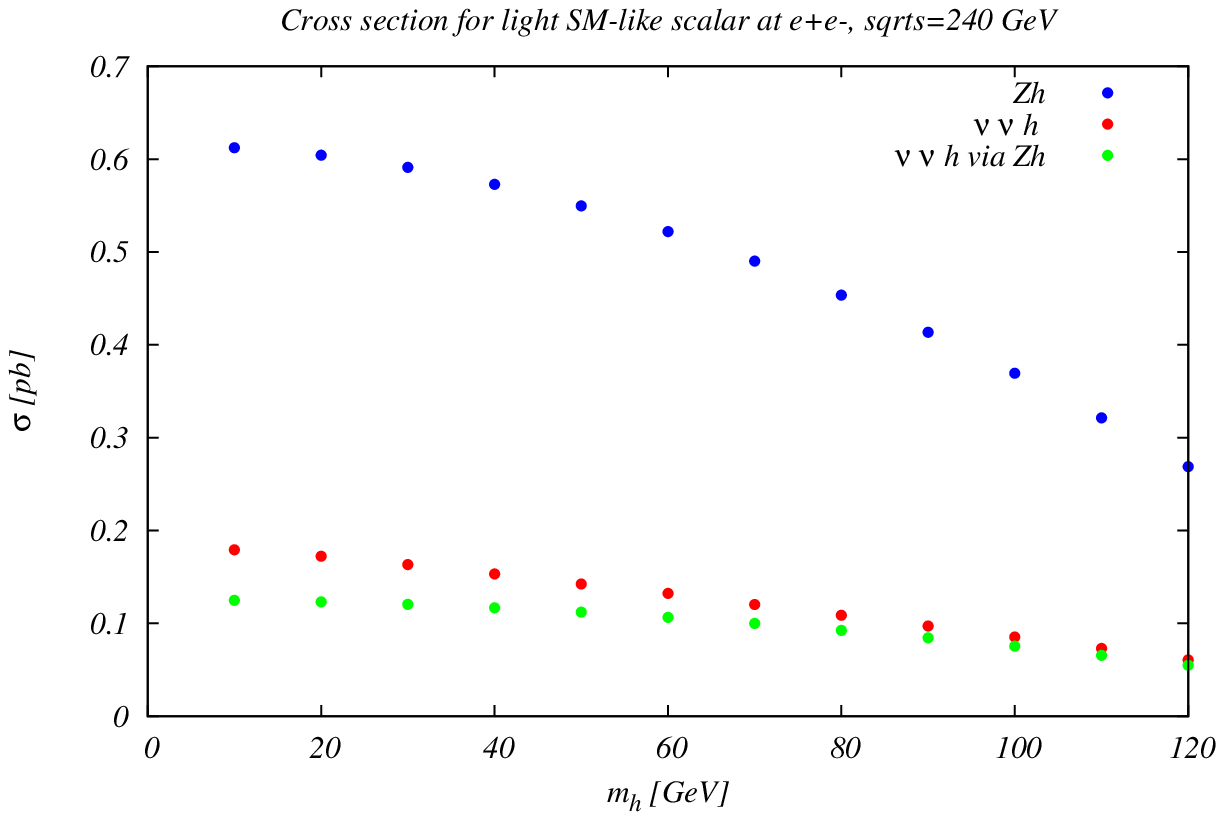}
\includegraphics[width=0.45\textwidth]{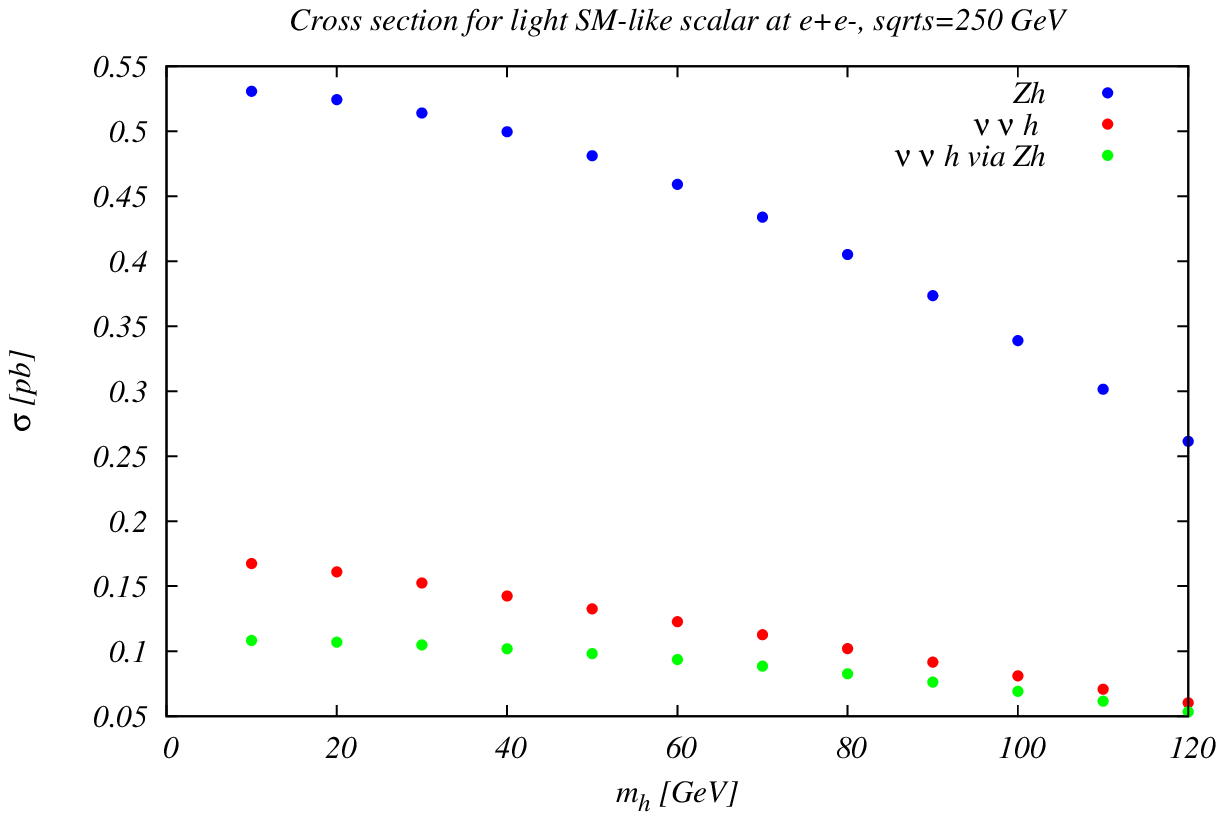}
\caption{\label{fig:prod250} Leading order production cross sections for $Z\,h$ and $h\,\nu_\ell\,\bar{\nu}_\ell$ production at an $e^+\,e^-$ collider with a com energy of 240 \GeV {\sl (left)} and 250 \GeV~ {\sl (right)} using Madgraph5 for an SM-like scalar h. Shown is also the contribution of $Z\,h$ to $\nu_\ell\,\bar{\nu}_\ell\,h$ using a factorized approach for the Z decay.}
\end{figure}
\end{center}

Figure \ref{fig:prod250} shows the production cross sections for these processes for a center-of-mass energy of $\,\sim\,240\,-\,250\,\GeV$. Using these predictions, and taking into account standard rescaling $\sim\,0.1$, around $10^5-10^6$ events could be produced with ILC, FCC-ee, and CEPC design luminosities \cite{Bambade:2019fyw,FCC:2018evy,CEPCStudyGroup:2018ghi}.

\subsection{Dedicated studies}
\subsubsection{Light scalars in $Zh$ production}
\begin{center}
\begin{figure}
\includegraphics[width=0.45\textwidth, angle=-90]{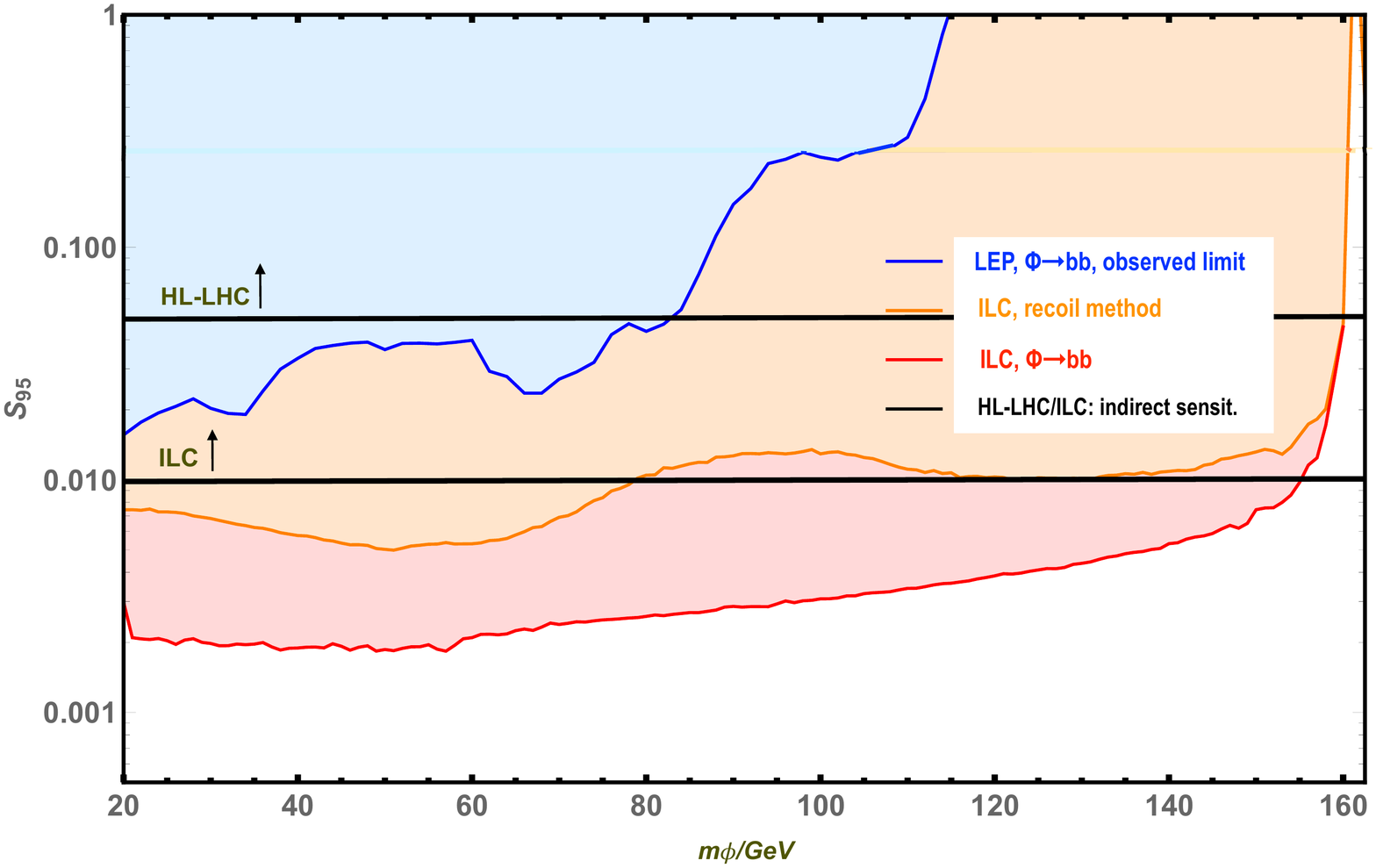}
\caption{\label{fig:lepgea} Sensitivity predictions for an ILC with a com energy of 250 \GeV~ from \cite{Drechsel:2018mgd}. See text for details.}
\end{figure}
\end{center}

\begin{center}
\begin{figure}
\includegraphics[width=0.45\textwidth]{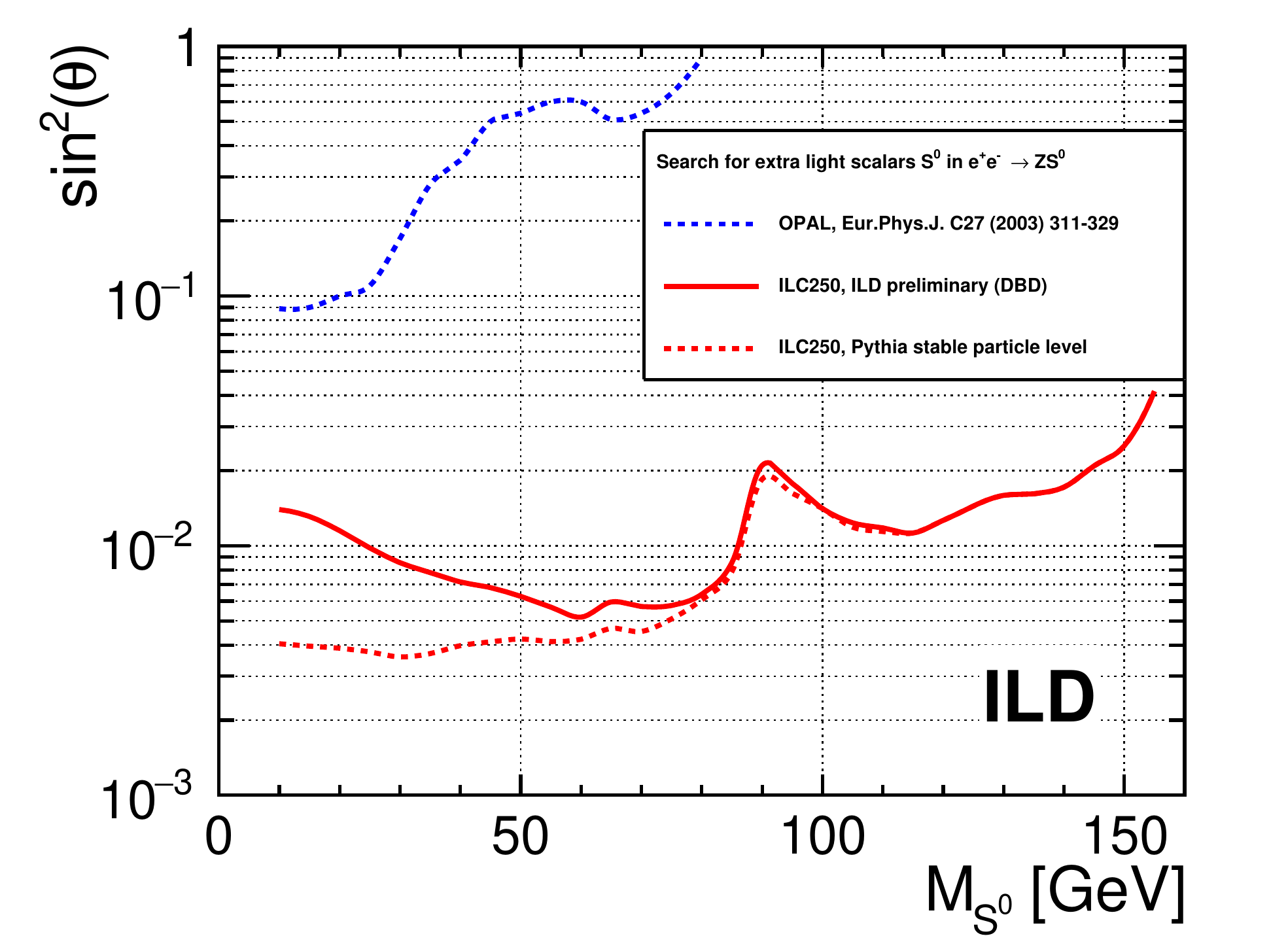}
\includegraphics[width=0.45\textwidth]{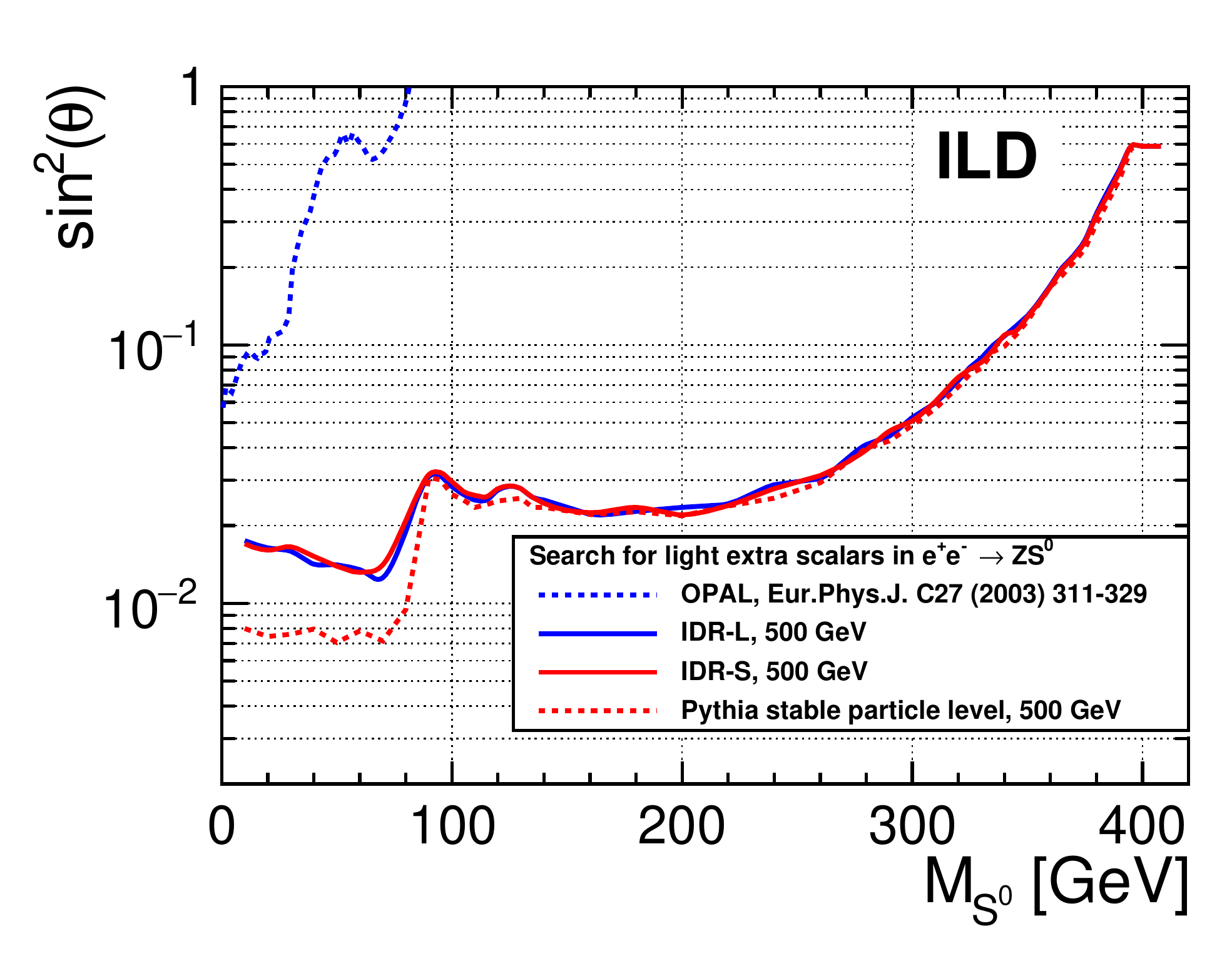}
\caption{\label{fig:jennyea} Upper bounds on the mixing angle for the model discussed in \cite{Wang:2020lkq}, in a comparison of different detector concepts and using the recoil method.}
\end{figure}
\end{center}

Not many dedicated studies exist that investigate low-mass scalars at Higgs factories. We here point to a study \cite{Drechsel:2018mgd} that investigates the sensitivity of the ILC for low-mass scalars in $Z\,h$ production, either using pure $Z$ recoil ("recoil method") or taking the light scalar decay into $b\,\bar{b}$ into account. The y-axis shows the $95\,\%$ CL limit for agreement with a background only hypothesis, which can directly be translated into an upper bound on rescaling. The authors validate their method by reproducing LEP results \cite{LEPWorkingGroupforHiggsbosonsearches:2003ing,ALEPH:2006tnd} for these channels prior to applying their method to the ILC. Their predictions are shown in figure \ref{fig:lepgea}.
A more detailed study along similar lines using the recoil method only and comparing different detector options has been presented recently in \cite{Wang:2020lkq}. We display their results in figure \ref{fig:jennyea}. The authors perform their analysis in a model where the coupling of the new resonance is rescaled by a mixing angle $\sin\theta$; therefore, their results can directly be compared with the ones presented in \cite{Drechsel:2018mgd} and figure \ref{fig:lepgea}.
\subsubsection{Higgsstrahlung and decay into two light scalars}
\begin{center}
\begin{figure}
\includegraphics[width=0.45\textwidth]{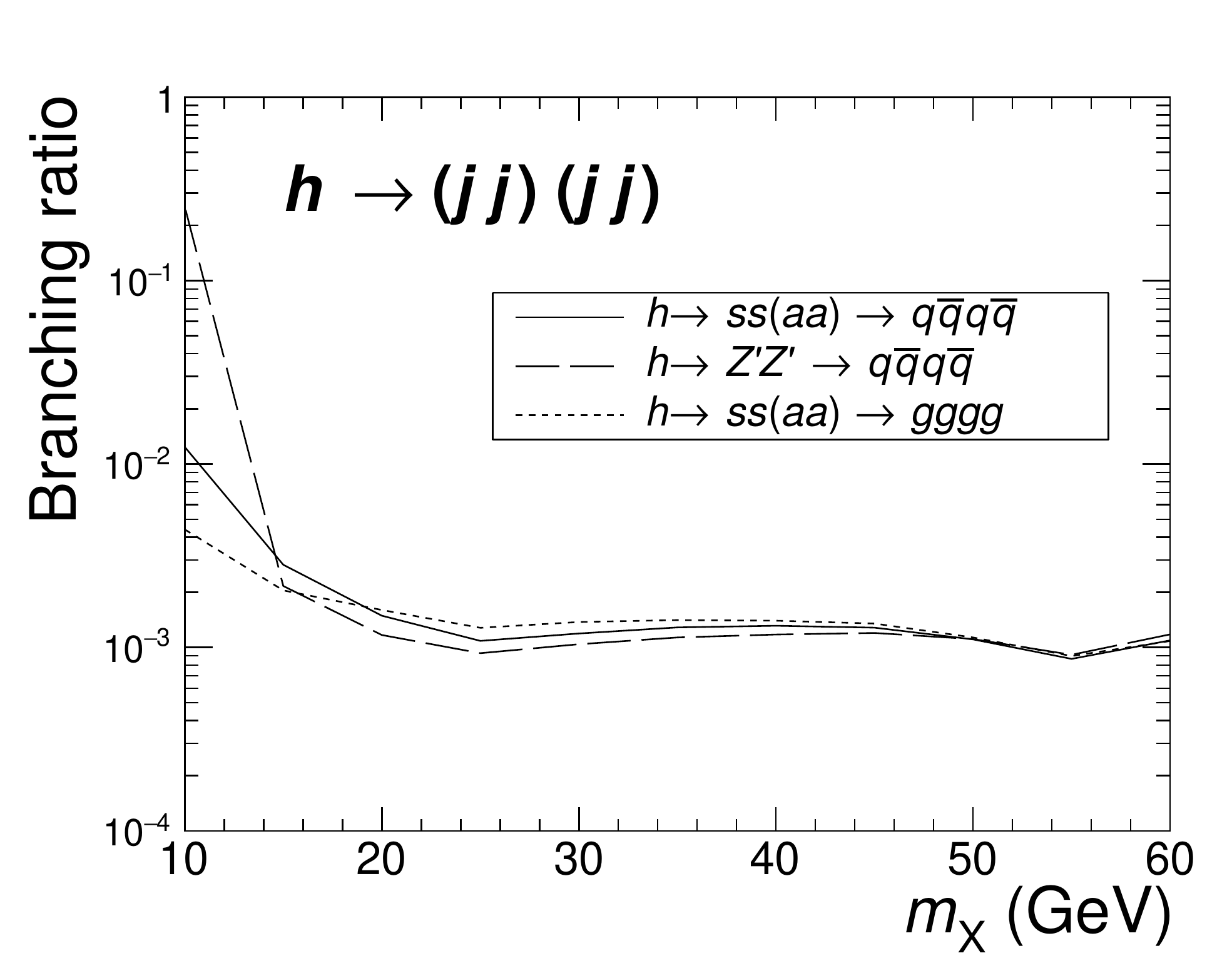}
\includegraphics[width=0.45\textwidth]{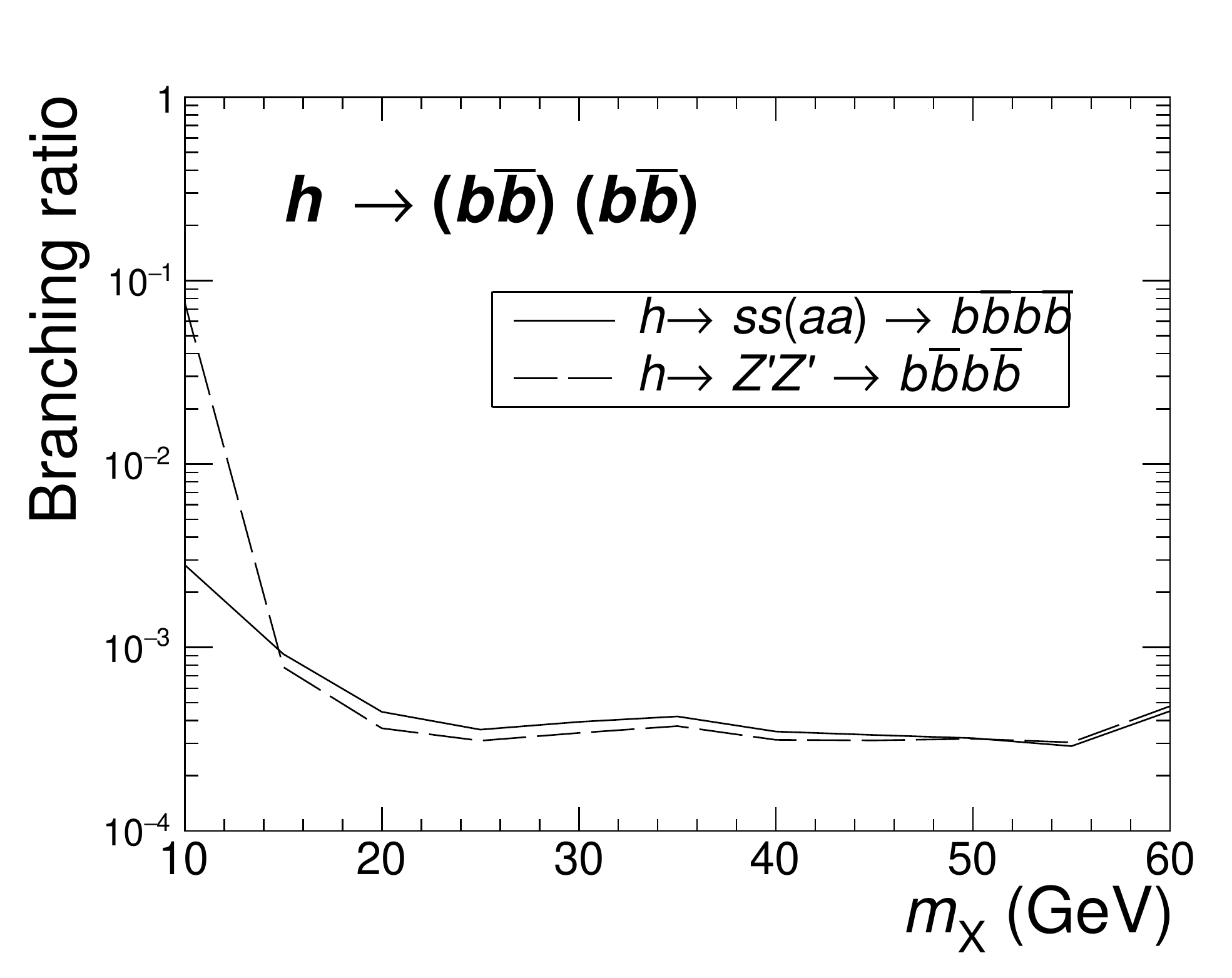}
\caption{\label{fig:discalar} 95 $\%$ confidence bounds on branching ratios for Higgs decay into a pair of lighter particles, for a com energy of 240 \GeV and $\int\mathcal{L}\,=\,5\,\ab^{-1}$. Taken from \cite{Liu:2016zki}. }
\end{figure}
\end{center}

\begin{center}
\begin{figure}
\includegraphics[width=0.5\textwidth]{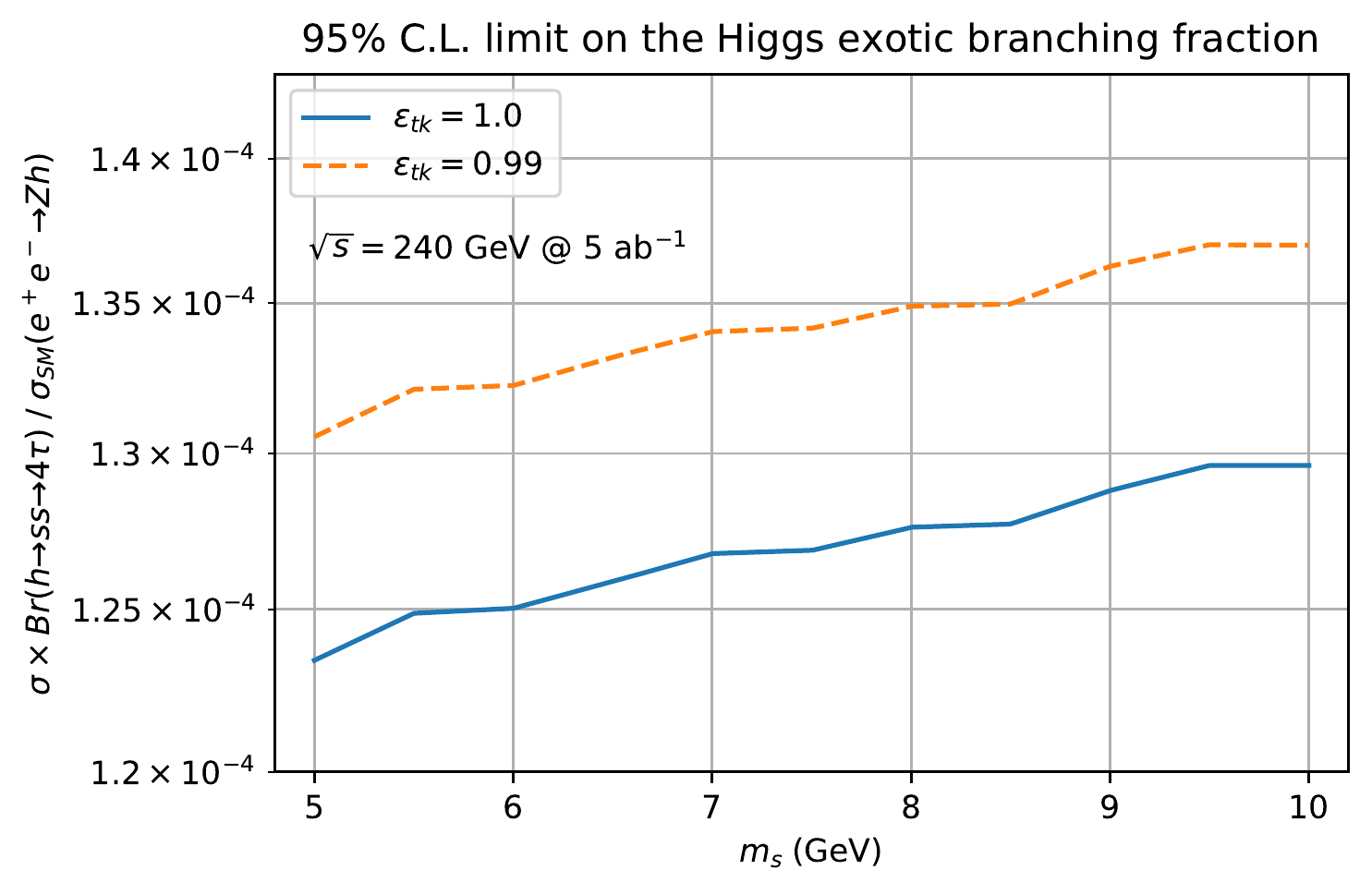}
\caption{\label{fig:ee4tau} Bounds on decay of the SM Higgs boson into two light scalars, with a 4 $\tau$ final state, at an $e^+e^-$ collider with a com energy of 240 \GeV, with different assumptions on tracking efficiencies. Taken from \cite{Shelton:2021xwo}.}
\end{figure}
\end{center}

In \cite{Liu:2016zki}, the authors consider Higgs-strahlung at a 240 \GeV $e^+e^-$ collider, where the Higgs subsequently decays into two light scalar states. The give 95 $\%$ confidence level bounds for the branching ratios into the decay productions of the two light scalars as a function of the light scalar masses for an integrated luminosity of $\int\mathcal{L}\,=\,5\,\ab^{-1}$ following a detailed study. Their results are subsequently used by many authors as standard reference. We show their results for various channels in figure \ref{fig:discalar}.
A more recent study \cite{Shelton:2021xwo} investigates the same process, but for $4\,\tau$ final states, for the same center-of-mass energy and integrated luminosity. The results, for varying values of tracking efficiency, are shown in figure \ref{fig:ee4tau}. Note that curent constraints on the invisible branching ratio of the Higgs, the signal strength, as well as SM-like decays of the light scalars currently render a bound $\lesssim\,10^{-3}$.

\begin{center}
\begin{figure}
\includegraphics[width=0.85\textwidth]{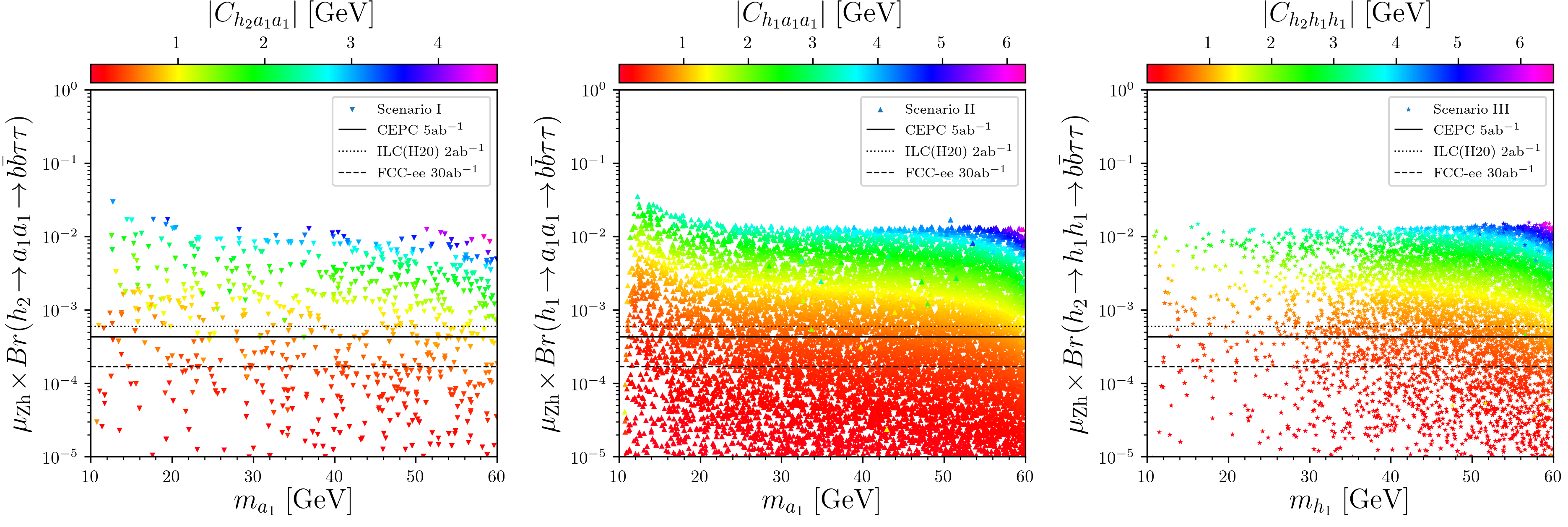}
\includegraphics[width=0.85\textwidth]{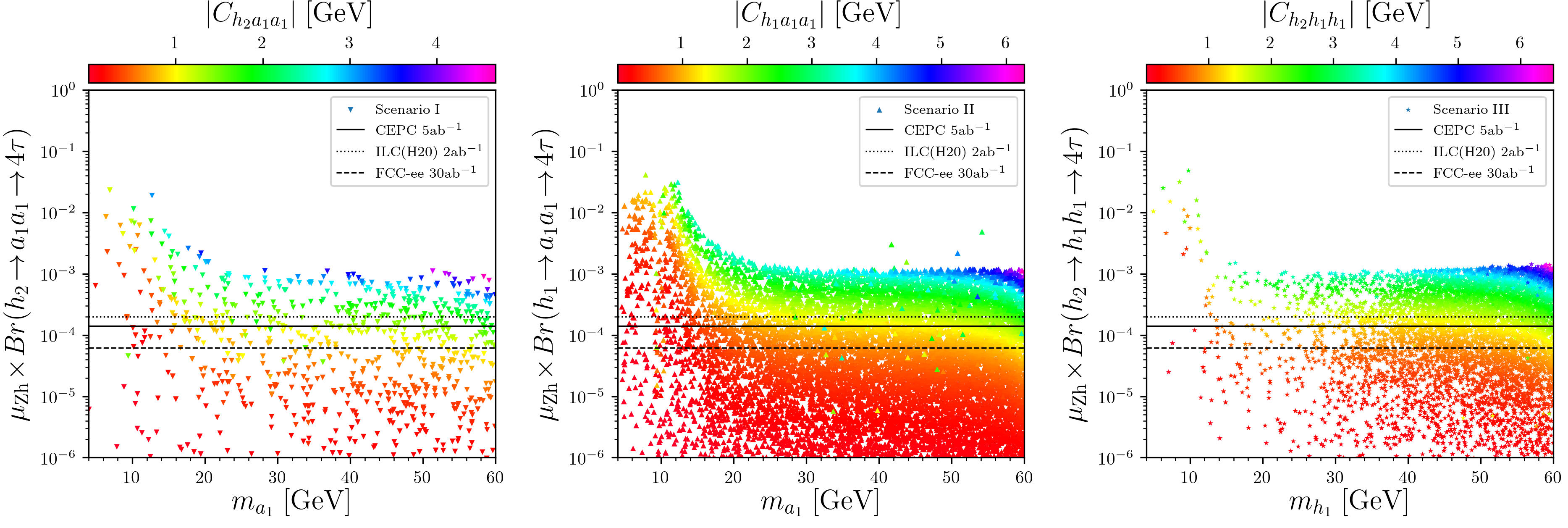}
\caption{\label{fig:scnmssm} Allowed rates for various Higgsstrahlung processes with successive decays into two light scalars. {\sl Top:} $2\,b\,2\,\tau$ final state. {\sl Bottom:} $4\tau$ final state. Also shown are expected upper bounds for various collider machines, with projections from \cite{Liu:2016zki}. Figure taken from \cite{Ma:2020mjz}.}
\end{figure}
\end{center}

Several works make use of the bounds derived in \cite{Liu:2016zki}. In \cite{Ma:2020mjz}, the authors investigate the allowed parameter space in the scNMSSM, an NMSSM extension that relaxes unification requirements at the GUT scale \cite{Das:2013ta,Ellwanger:2014dfa,Nakamura:2015sya,Wang:2018vxp}, also known as NUHM, which contains in total 5 scalar particles; if CP is conserved, 3 are CP-even and 2 CP-odd. The authors investigate various bounds on the models parameter space, and show the allowed scan points and predictions for the above channels for various scalar combinations. We show their results in figure \ref{fig:scnmssm}.

\begin{center}
\begin{figure}
\includegraphics[width=0.65\textwidth]{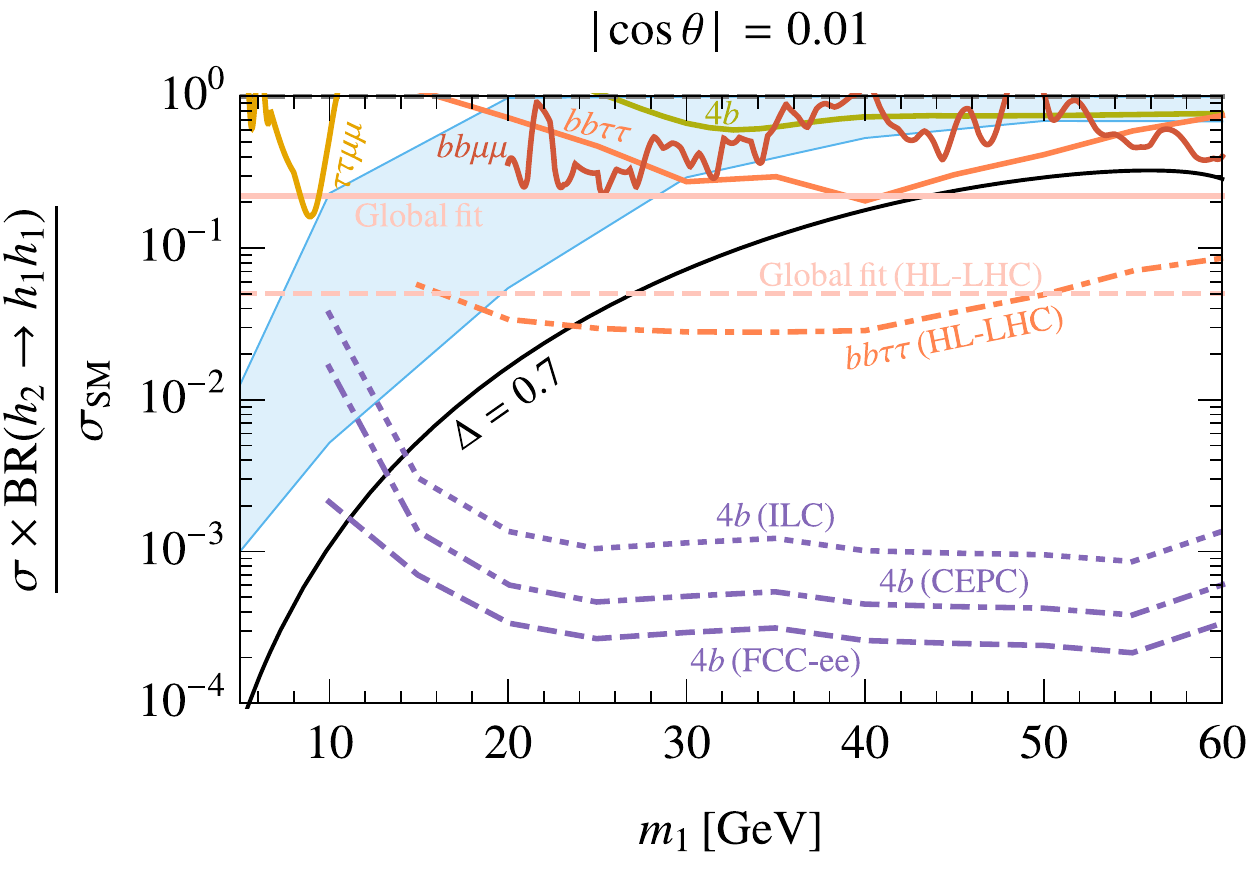}
\caption{\label{fig:rmewp} Expected bounds on Higgs production via Higgs strahlung and subsequent decay into two light scalars, in the singlet extension scenario discussed in \cite{Kozaczuk:2019pet}. The blue band denotes the region where a strong first-order electroweak phase transition is possible. We see that $e^+e^-$ Higgs factories are required on order to confirm or exclude such scenarios. Figure taken from \cite{Kozaczuk:2019pet}.}
\end{figure}
\end{center}

Finally, in simple singlet extensions it is possible to test regions in the models parameter space which can lead to a strong first-order electroweak phase transition. Several authors have worked on this; we here show results from \cite{Kozaczuk:2019pet}, where in addition several collider sensitivity projections are shown, including the bounds derived in \cite{Liu:2016zki}. From figure \ref{fig:rmewp}, it becomes obvious that $e^+e^-$ Higgs factories would be an ideal environment to confirm or rule out such scenarios.
Related work, with a spontaneous breaking of the imposed $\mathbb{Z}_2$ symmetry, has been presented in \cite{Carena:2019une}.
\subsubsection{Other channels}
\begin{center}
\begin{figure}
\includegraphics[width=0.45\textwidth]{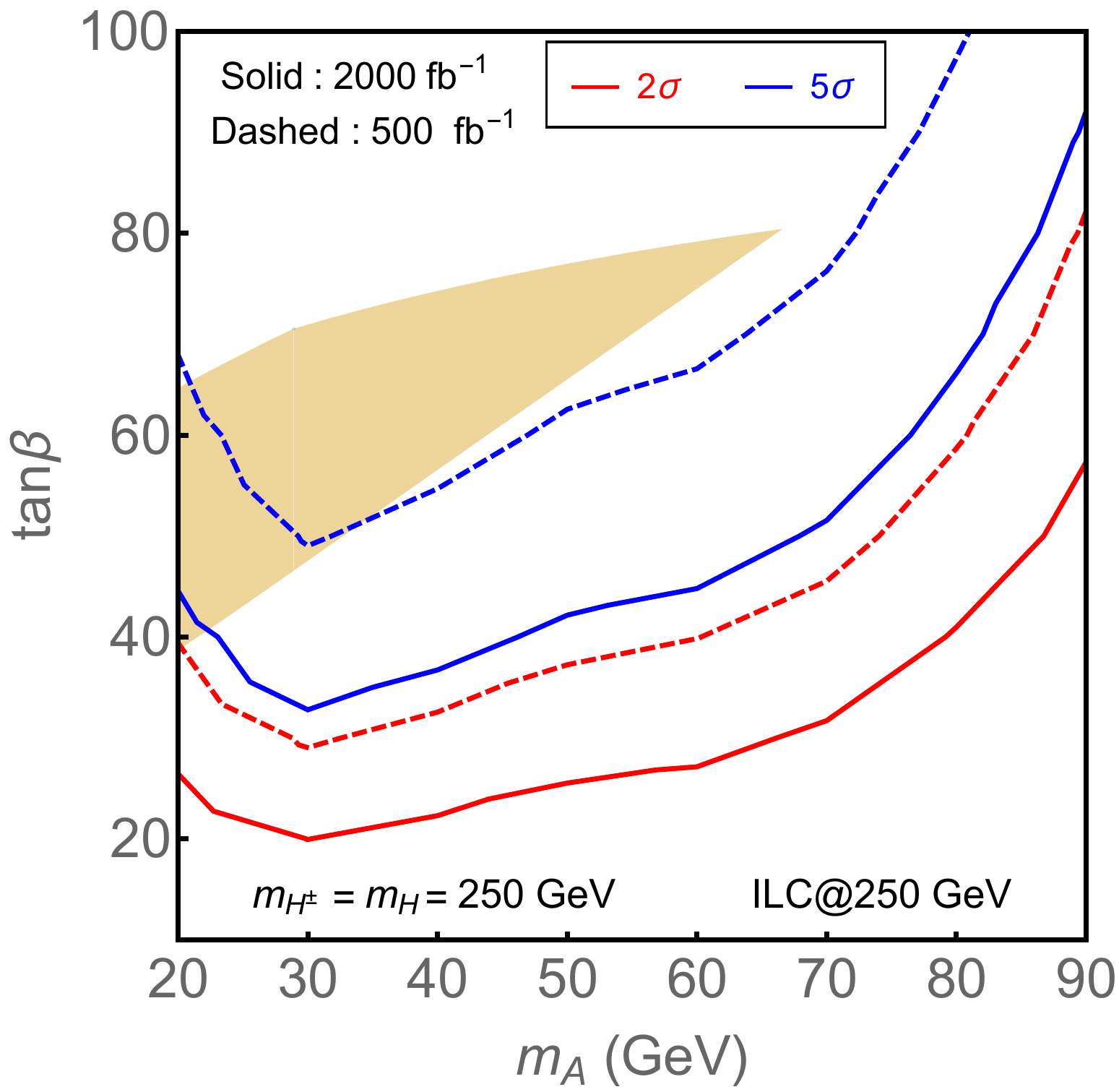}
\caption{\label{fig:astr} Exclusion and discovery regions in the 2HDM type X model, in the $m_A,\,\tan\be$ plane. The color region additionally explains the current $g_\mu-2$ discrepancy. Taken from \cite{Chun:2019sjo}.}
\end{figure}
\end{center}

In \cite{Chun:2019sjo}, the authors investigate a slightly different channel, i.e. tau pair-production where a light pseudoscalar is radiated off one on the outgoing fermion lines and decays again into $\tau^+\tau^-$, leading to a 4 $\tau$ final state. They are investigating this within a type X 2HDM, which in addition allows them to explain the current discrepacy between theoretical prediction and experiment for the anomalous magnetic momentum of the muon. They perform a detailed study including background and determine 2 and 5 $\sigma$ countours in the $m_A,\,\tan\be$ plane, where $\tan\be$ denotes the ratio of the vevs of the two doublets. Their results are shown in figure \ref{fig:astr}.

\begin{center}
\begin{figure}
\includegraphics[width=0.45\textwidth]{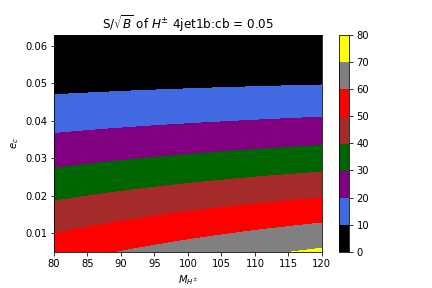}
\includegraphics[width=0.45\textwidth]{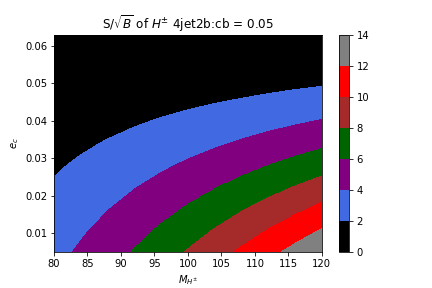}
\caption{\label{fig:charged} Significances as a function of charged scalar mass and charm tagging efficiency at an 240 \GeV~ CEPC, at an integrated luminosity of $1\,\ab^{-1}$, within a 3HDM as presented in \cite{Akeroyd:2019mvt}, considering a $c\bar{c}b\bar{b}$ final state. Figures taken from that reference.}
\end{figure}
\end{center}

It is also interesting to investigate models with give the possibility of light charged scalars. A corresponding study has been performed in \cite{Akeroyd:2019mvt}, where the authors consider charged scalar pair-production within a 3HDM, with successive decays into $c\,\bar{c}\,b\,\bar{b}$ final states. The authors perform a detailed study and present their results in the 1 and 2 b-jet tagged category, as a function of light scalar mass and charm tagging efficiency. We show the corresponding significances in figure \ref{fig:charged}, for a com energy of 240 \GeV and an integrated luminosity of $1\,\ab^{-1}$.

\subsection{Cross section predictions}
\begin{center}
\begin{figure}
\includegraphics[width=0.49\textwidth]{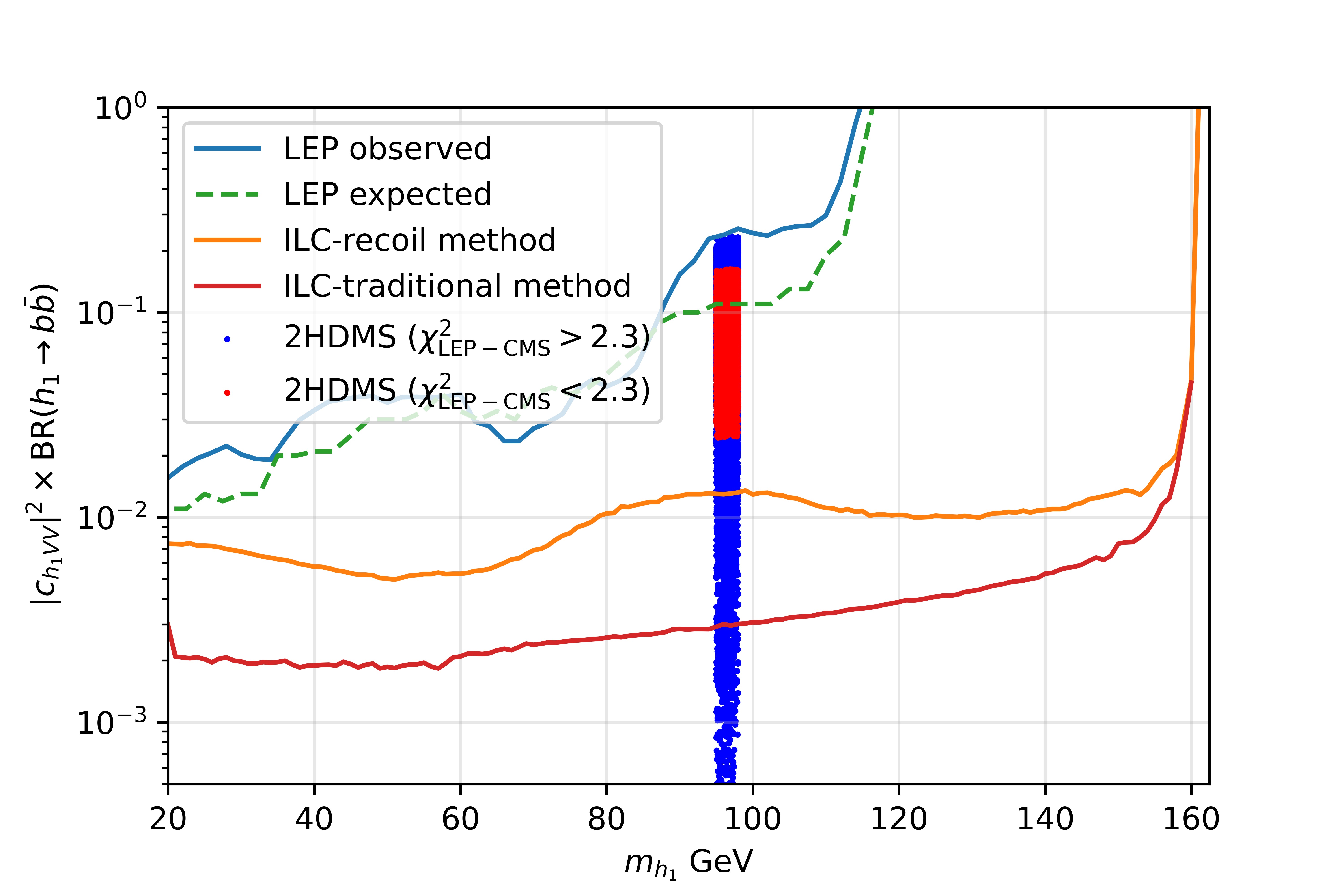}
\includegraphics[width=0.49\textwidth]{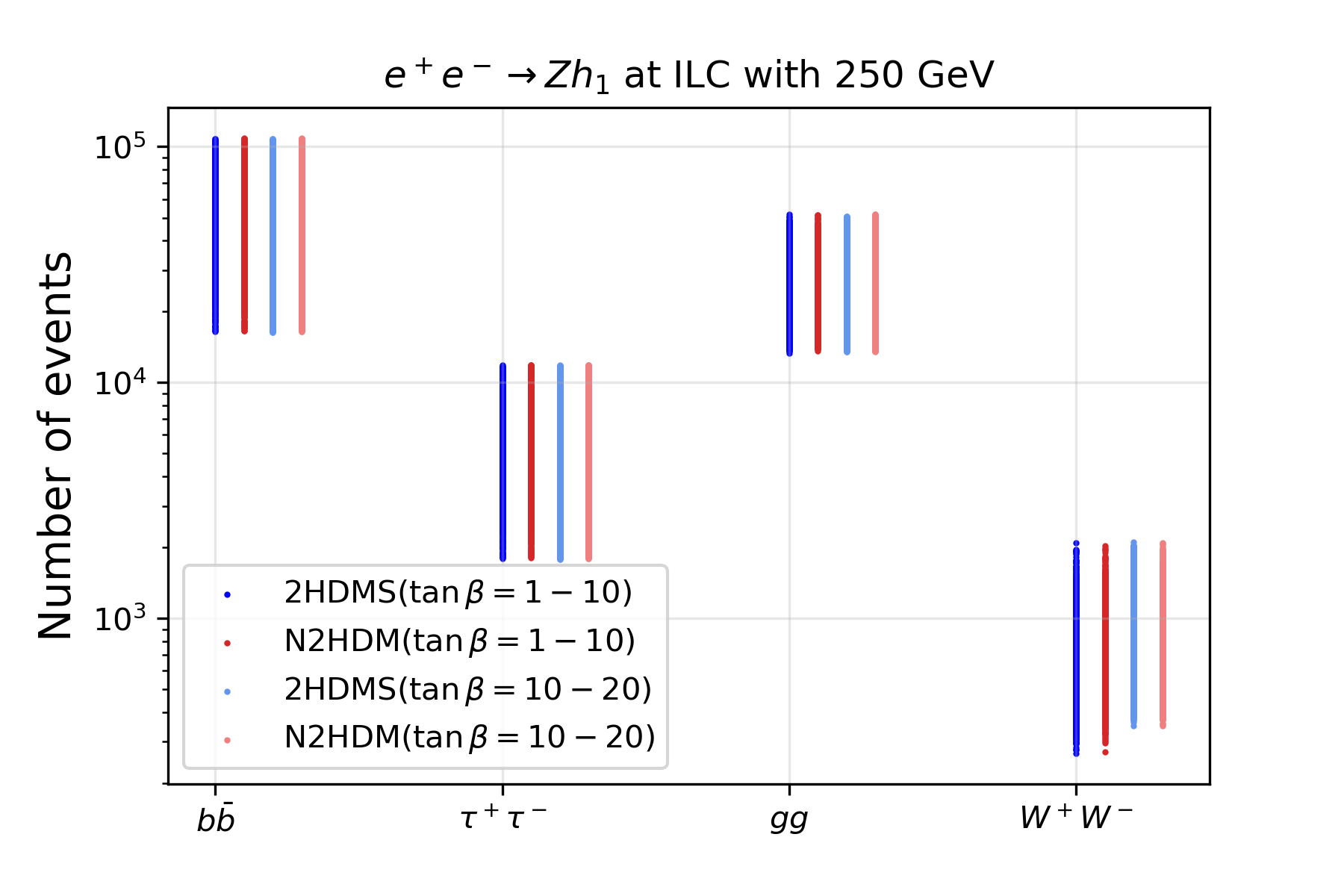}
\caption{\label{fig:svenea} {\sl Left:} Points in the 2HDMs that agree with both CMS and LEP excess and which can be probed at the ILC. {\sl Right:} predicted rates in the 2HDMS and N2HDM at 250 \GeV using full target luminosity.}
\end{figure}
\end{center}

Inspired by possible low-mass excesses in at LEP \cite{ALEPH:2006tnd} and CMS \cite{CMS:2018cyk}, in \cite{Heinemeyer:2021msz} several models are fitted to these excesses that contain singlet and doublet extensions of the SM scalar sector; in particular, they consider models with an additional doublet as well as a (complex) singlet, labelled N2HDM and 2HDMs, respectively. For both models, as well as varying $\tan\be$ ranges (where $\tan\be$ denotes the ratio of the vevs in the 2HDM part of the models), the authors consider the possibility to explain the observed accesses and give rate predictions for a $250\,\GeV$ collider with a total luminosity of $\mathcal{L}\,=\,2\,\ab^{-1}$. We display their results in figure \ref{fig:svenea}. We see that also other final states for the $h$ decay, as e.g. $\tau^+\tau^-,\,gg,$ or $W^+W^-$ can render sizeable rates.
\section{Other center of mass energies}
\begin{center}
\begin{figure}
\includegraphics[width=0.45\textwidth]{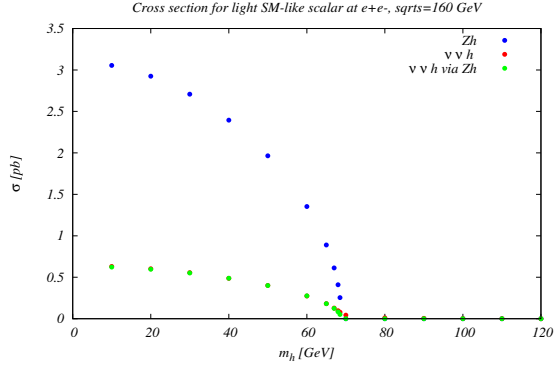}
\caption{\label{fig:prod160} As figure \ref{fig:prod160}, for a com of 160 \GeV. We assume onshell final states.}
\end{figure}
\end{center}

The FCC-ee and CEPC colliders are supposed to also run with a center-of-mass energy of $\sim\,160\,\GeV$, already tested at LEP. In analogy to figure \ref{fig:prod250}, in figure \ref{fig:prod160} we show again cross section predictions for the process $Z\,h$ in dependence of the mass of $h$, assuming a SM-like scalar. Note we here assume onshell production of $Z\,h$, which leads to a hard cutoff for $M_h\,\sim\,70\,\GeV$. Detailed studies should in turn assume contributions from offshell $Z$s and $h$s as well. 
We see that for this lower com energy, there is basically no contribution to the $\nu_\ell\,\bar{\nu}_\ell h$ final state that does not originate from $Z\,h$. Using FCC-ee target luminosity for this energy, and again assuming a general suppression factor $\,\sim\,0.1$ stemming from signal strength, we expect up to $10^6$ events depending on the mass of the additional scalar.

For this center-of-mass energy, several searches exist which have already been performed at LEP and are summarized in \cite{OPAL:2002ifx,ALEPH:2006tnd}, concentrating on $Z\,h$, $h_1\,h_2,$ and $h_1\,h_1\,h_1$ final states, which could be further pursued in future collider studies. We want to note that the luminosity at FCC-ee at this center of mass energy is exceeding LEP luminosity by several orders of magnitude.

\begin{center}
\begin{figure}
\includegraphics[width=0.45\textwidth]{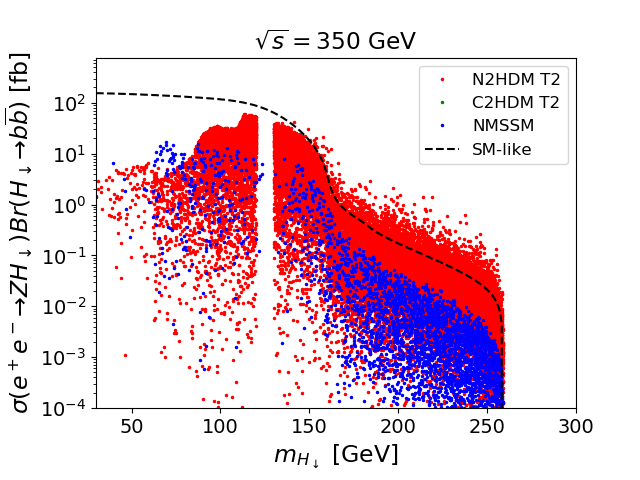}
\includegraphics[width=0.45\textwidth]{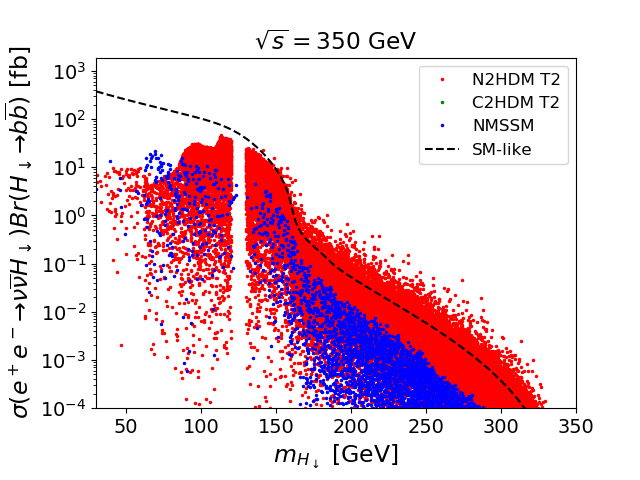}
\caption{\label{fig:350} Achievable rates for various light scalar production modes at an $e^+e^-$ collider with a com energy of 350 \GeV, in various 2HDM variant models. Figure taken from \cite{Azevedo:2018llq}.}
\end{figure}
\end{center}

Finally, we present a study that investigates various types of 2HDMs containing several neutral scalars \cite{Azevedo:2018llq}, for a collider energy of 350 \GeV. The authors perform a scan of the allowed parameter space and render predictions for the Higgs-strahlung process as well as $\nu_\ell\,\bar{\nu}_{\ell} h$ final states with the scalar decaying into $b\,\bar{b}$ pairs. We show their results in figure \ref{fig:350}.

\section{Conclusion}
In this short note, I have presented several models and searches that investigate the sensitivity of future $e^+e^-$ machines for scalars with masses $\lesssim\,125\,\GeV$. This corresponds to a summary of several talks I have recently given and is thereby not meant to be inclusive. I have pointed to models that allow for such light scalars, as well as several references that either provide rates or pursue dedicated studies. I have also pointed to the connection of low-scalar searches at such colliders and the electroweak phase transition within certain models. My impression is that further detailed studies are called for, with a possible focus on so-called Higgs factories with center-of-mass energies around 240-250 \GeV.
\section*{Acknowledgements}
I thank Sven Heinemeyer and the conveners of the CEPC workshop for inspiring me to set up an overview on these models. Several authors of the references cited here were also helpful in answering specific questions regarding their work.

\end{document}